\documentclass{ieeeaccess}
\usepackage{colortbl}
\usepackage{cite}
\usepackage{amsmath,amssymb,amsfonts}
\usepackage{algorithmic}
\usepackage{graphicx}
\usepackage{textcomp}
\usepackage{braket} 
\usepackage{makecell}
\def\BibTeX{{\rm B\kern-.05em{\sc i\kern-.025em b}\kern-.08em
    T\kern-.1667em\lower.7ex\hbox{E}\kern-.125emX}}
\begin{document}
\history{Date of publication xxxx 00, 0000, date of current version xxxx 00, 0000.}
\doi{10.1109/TQE.2020.DOI}

\title{Quantum-inspired Hash Function Based on Parity-dependent Quantum Walks with Memory (August 2023)}
\author{\uppercase{Qing Zhou}\authorrefmark{1}, \uppercase{Xueming Tang}\authorrefmark{1}, \uppercase{Songfeng Lu}\authorrefmark{1}, and \uppercase{Hao Yang}\authorrefmark{1}}
\address[1]{Hubei Key Laboratory of Distributed System Security, Hubei Engineering Research Center on Big Data Security, School of Cyber Science and Engineering, Huazhong University of Science and Technology, Wuhan, 430074, China}
\tfootnote{This work was supported in part by the National Natural Science Foundation of China under Grant 62101197, in part by the China Postdoctoral Science Foundation under Grant 2021M691148.}

\markboth
{Zhou \headeretal: Quantum-inspired Hash Function Based on Parity-dependent Quantum Walks with Memory}
{Zhou \headeretal: Quantum-inspired Hash Function Based on Parity-dependent Quantum Walks with Memory}

\corresp{Corresponding author: Xueming Tang (email: xmtang@hust.edu.cn).}

\begin{abstract}
    In this paper, we develop a generic controlled alternate quantum walk model (called CQWM-P) by combining parity-dependent quantum walks with distinct arbitrary memory lengths and then construct a quantum-inspired hash function (called QHFM-P) based on this model. Numerical simulation shows that QHFM-P has near-ideal statistical performance and is on a par with the state-of-the-art hash functions based on discrete quantum walks in terms of sensitivity of hash value to message, diffusion and confusion properties, uniform distribution property, and collision resistance property. Stability test illustrates that the statistical properties of the proposed hash function are robust with respect to the coin parameters, and theoretical analysis indicates that QHFM-P has the same computational complexity as that of its peers. 
\end{abstract}


\titlepgskip=-15pt

\maketitle

\section{Introduction}
\label{sec:1}
Cryptographic hash function acts as a key component of identification, message authentication, digital signatures, and pseudorandom number generation. From a security perspective, cryptographic hash functions can be divided into two broad categories: provably secure hash functions based on hard mathematical problems and dedicated hash functions based on ad-hoc constructions, especially iterative constructions of one-way compression functions. The former only satisfy computational security and are inefficient to be used in practice; the latter are efficiently computable, but the security of which is not built on a firm foundation.

To develop a secure and efficient hash function, more and more researchers have shown interests in hash functions based on quantum computing~\cite{Zhou2021Hash,Hou2023Hash,Shi2022Hash,ShiJ2022Hash,Yang2021qwHash,Yang2019qwHash,Yang2018qwHash264,Yang18qwHash221,Li2018qwHash,Cao2018qwHash,Yang2016qHash,Li2013qwHash}, especially on discrete quantum walks~\cite{Zhou2021Hash,Hou2023Hash,Yang2021qwHash,Yang2019qwHash,Yang2018qwHash264,Yang18qwHash221,Li2018qwHash,Cao2018qwHash,Yang2016qHash,Li2013qwHash} (hereafter, simply DQW-based hash functions). The security of this kind of hash functions is based on quantum mechanics; more precisely, it is ensured by the theoretically infinite possibilities of the initial state and the irreversibility of modulo operation~\cite{Yang2018qwHash264}. The output results of DQW-based hash functions are more attainable by classical simulation of the quantum walk process, where the final amplitudes can be calculated with linear overhead with respect to the number of steps. For this reason, a DQW-based hash function can be considered as a “quantum inspired” algorithm, where everything is calculated classically using some desired properties of a quantum system and its dynamics. In what follows, hash schemes constructed by simulating discrete quantum walks and then calculating hash values from the resulting probability distributions are called DQW-based hash functions or quantum inspired hash functions interchangeably.

Quantum inspired hash functions was first explored by Li et al. in~\cite{Li2013qwHash}, and further intensively discussed by Cao et al. in~\cite{Cao2018qwHash}, Yang et al. in~\cite{Yang2021qwHash,Yang2019qwHash,Yang2018qwHash264,Yang18qwHash221,Yang2016qHash}, and zhou et al. in~\cite{Zhou2021Hash}. Until Zhou et al.'s scheme (called QHFM)~\cite{Zhou2021Hash}, all quantum inspired hash functions simply use quantum walks with different coin parameters, where only the coin operator is controlled by the message bit. QHFM is based on quantum walks with memory, where an additional changeable transform---the direction-determine operator is also controlled by the message bit. More recently, Hou et al.~\cite{Hou2023Hash} has proposed a hash function QHFL based on lively quantum walks, where the shift operator contains a liveliness parameter, whose value (one out of two) is determined by the message bit. The good statistical performance of QHFM and QHFL suggests that combining quantum walks differing in component operators other than coin transform is a promising way to construct good hash functions.

Quantum walks with memory (QWM)~\cite{Gettrick2010QW1M,Zhou2019QW2M,Li2020QWM,Li2020szegedy,Dai2020QWM,Dai2018QWM,Molfetta2018QWM,Li2016QWM,Gettrick2014QWM,Konno2010QWM} are types of modified quantum walks where the next direction of the walking particle is governed by the direction-determine operator, which specifies how the latest path together with the coin state affects the movement direction of the particle; different QWM models have different direction-determine operators. Unlike usual quantum walks without memory~\cite{Ambainis2001QW} or lively quantum walks~\cite{Hou2023Hash}, where the changeable operator (the coin or shift operator) can be characterized by a single parameter, the direction-determine operator of QWM is highly flexible. It is influenced by two factors: the memory length and the movement rule, the latter can be an arbitrary relation between the movement direction and the recorded shifts of the walker, which cannot simply be dictated by a few parameters. Hence, QWM have great potential to be used to construct various quantum-inspired hash functions. 

Here, we present an alternative QWM-based hash function QHFM-P, which is inspired by quantum walks with memory provided by parity of memory~\cite{Li2020QWM}. The proposed hash function is on a par with the state-of-the-art DQW-based hash functions in terms of sensitivity of hash value to message, diffusion and confusion properties, uniform distribution property, collision resistance property, and computational complexity. Furthermore, it can preserve near-ideal statistical performance when changing the values of its coin parameters, indicating that QHFM-P has nice stability with respect to different coin angles.
\section{Controlled quantum walk with memory depending on the parity of memory}\label{sec:2}
One-dimensional quantum walks with $m$-step memory depending on the parity of memory (QWM-P), or parity-dependent quantum walks with $m$-step memory on the line~\cite{Li2020QWM}, is a quantum system living in a Hilbert space $\mathcal{H}_p\otimes\mathcal{H}_{d_m}\otimes\cdots\otimes\mathcal{H}_{d_2}\otimes\mathcal{H}_{d_1}\otimes\mathcal{H}_c$ spanned by orthogonal basis states $\{\Ket{x,d_m,\dots,d_2,d_1,c}\vert d_m,\dots,d_1,c\in\mathbb{Z}_2;x\in\mathbb{Z}\}$, where $c$ is the coin state, $d_j$ (0 stands for left and 1 stands for right) records the shift of the walker $j$ steps before ($d_1$ is the direction of the most recent step, and $d_m$ is the earliest direction that is memorized), and $x$ is the current position of the walker. If the walker moves on a cycle with $n$ nodes, then $x\in\mathbb{Z}_n$. The one-step evolution of QWM-P may be decomposed into three parts: the first is a $2\times 2$ coin operator $C$ on subspace $\mathcal{H}_c$, here $C$ is parameterized by an angle $\theta$, i.e.,
\begin{equation}
    C=
    \begin{pmatrix}\label{eq:1}
    \text{cos}\theta & \text{sin}\theta \\
    \text{sin}\theta & -\text{cos}\theta
    \end{pmatrix};
\end{equation}
the second is the direction-determine transform $D$ on $\mathcal{H}_{d_m}\otimes\cdots\otimes\mathcal{H}_{d_2}\otimes\mathcal{H}_{d_1}\otimes\mathcal{H}_{c}$, whose action can be written as
\begin{equation}\label{eq:2}
\begin{split}
    \hat{D}:&\Ket{d_m, d_{m-1}, \dots ,d_2, d_1,c}\rightarrow\\
      &\Ket{d_{m-1},d_{m-2},\dots ,d_1, c\oplus 1\oplus ifeven(d_m,\dots ,d_1),c},
\end{split}    
\end{equation}
where $ ifeven(d_m,\dots ,d_1)$ equals 1 (respectively, 0) if the number of zeros in the memorized directions $ d_m,\dots ,d_1$ is even (respectively, odd); and the third is the shift operator $S$ on $\mathcal{H}_p\otimes\mathcal{H}_{d_1}$, whose action can be expressed as
\begin{equation}\label{eq:3}
    S:\Ket{x,d_1}\rightarrow\Ket{x+2d_1-1,d_1}.
\end{equation}
If the walker moves on a cycle with $n$ nodes, then the shift operator becomes $\Ket{x,d_1}\rightarrow\Ket{x+2d_1-1 (\bmod n),d_1}$.

A controlled one-dimensional quantum walks with memory depending on the parity of memory (CQWM-P) can be obtained by alternately applying QWM-P with different memory lengths (as well as different coin parameters). More precisely, CQWM-P evolves according to a $t$-bit binary string $msg=(m_1,m_2,\dots ,m_t)\in\{0,1\}^t$: at the $j$th time step, if $m_j=0$ ($j\in\{1,2,\dots, t$), then the walker performs QWM-P with $s_0$ step memory (denoted by QW$s_0$M-P) and with coin parameter $\theta_0$; if $m_j=1$, then the walker performs QWM-P with $s_1$ ($s_1\neq s_0$) step memory (denoted by QW$s_1$M-P) and with coin parameter $\theta_1$.

To enable QW$s_0$M-P and QW$s_1$M-P to be performed alternately, $|s_0-s_1|$ redundant qubits can be added to the walk process with less memory length, so that two walks live in the same Hilbert space. For instance, if $0<s_0<s_1$, then $s_1-s_0$ redundant qubits are added to QWM$s_0$P. In this case, the basis states of QWM$s_0$P becomes $\{\Ket{x,d_{s_1},\dots,d_2,d_1,c}|d_{s_1},\dots,d_1,d\in \mathbb{Z}_2;x\in\mathbb{Z}\}$, wherein the first $s_1-s_0$ qubits are invariant under the transforms of QWM$s_0$P, and the $D$ transform becomes
\begin{equation}\label{eq:4}
    \begin{split}
    D: \Ket{d_{s_1},\dots ,d_{s_0+1},d_{s_0},d_{s_0-1},\dots ,d_2,d_1,c}\rightarrow \\
    \Ket{d_{s_1},\!\dots\!,d_{s_0\!+\!1},d_{s_0\!-\!1},\!\dots\!,d_1,c\!\oplus\!1\!\oplus\!ifeven(d_{s_0},\dots ,d_1),c}. 
    \end{split}
\end{equation}    

In this work ,we focus on CQWM-P with one- and two-step memory, whose evolution operator controlled by $msg$ is the product of $t$ unitary transforms
\begin{equation}\label{eq:5}
    U_{msg}=U^{(m_t)}U^{(m_{t-1})}\cdots U^{(m_2)}U^{(m_1)},
\end{equation}
where $U^{(m_j)}$ is  the one-step transform defined as
\begin{equation}\label{eq:6}
    U^{(m_j)}=S\cdot \left(I_n\otimes D^{(m_j)}\right)\cdot \left(I_{4n}\otimes C^{(m_j)}\right).
\end{equation}

In Eq.~(\ref{eq:6}), $C^{(0)}$ and $C^{(1)}$ are coin operators parameterized by $\theta_0$ and $\theta_1$, respectively; $I_{4n}$ and $I_n$ are $4n\times 4n$ and $n\times n$ identity operators, respectively; $S$ is the conditional shift operator controlled by the next direction $d_1$, such a direction is determined by an $8\times 8$ unitary operator $D^{(m_j)}$. If $m_j=0$, then $D^{(m_j)}$ is the direction-determine transform of QW1M-P, i.e., $\hat{D}^{(0)}:\Ket{d_1,c}\rightarrow\Ket{c\oplus 1\oplus d_1,c}$; and if $m_j=1$, then $D^{(m_j)}$ is the  direction-determine transform of QW2M-P, i.e., $D^{(1)}:\Ket{d_2,d_1,c}\rightarrow \Ket{d_1,c\oplus (d_1\oplus d_2),c}$. By appending a redundant qubit $d_2$ to the state of QW1M-P and letting $D^{(0)}:\Ket{d_2,d_1,c}\rightarrow\Ket{d_2,c\oplus 1\oplus d_1,c}$ determines the next direction if the controlling bit equals 0, the walk process can switch freely between QW1M-P and QW2M-P.
\section{Hash Function Using Quantum Walks with One- and Two-step Memory on Cycles}\label{sec:3}
The proposed hash function is constructed by numerically simulating CQWM-P with one- and two-step memory on cycles under the control of the input message, and then calculating the hash value from the resulting probability distribution of this walk.

Specifically, our hashing algorithm is parameterized by three positive integers $\{n,m,l|n \bmod 2=1,10^l\gg 2^m\}$ and three angles $\{\theta_0,\theta_1,\alpha|\theta_0,\theta_1,\alpha\in(0,\pi/2)\}$, where $n$ specifies the total number of nodes in the cycle that the walker moves along, $m$ is the number of hash bits that are contributed by each node, $l$ is the number of digits in the probability value (associated with each node) that are used to calculate the hash result, $\theta_0$ (respectively, $\theta_1$) is the coin parameter of QW1M-P (respectively, QW2M-P), and $\alpha$ is the parameter of the initial state of the walker. Given the input message $msg$, the $m\times n$-bit hash value $H(msg)$ is calculated as follows.
\begin{enumerate}
    \item[1)] Initialize the walker in the state $\Ket{\psi_0}\!=\!\text{cos}\alpha\Ket{0,\!1,\!0,\!0}+\text{sin}\alpha\Ket{0,1,0,1}$;
    \item[2)] Apply $U_{msg}$ to $\Ket{\psi_0}$ and generate the resulting probability distribution $prob=(p_0,p_1,\dots,p_{n-1})$, where $p_x$ is the probability that the walker locates at node $x$ when the walk is finished.
    \item[3)] The hash value of $msg$ is a sequence of $n$ blocks $H(msg)=B_0\lVert B_1\lVert \dots \lVert B_{n-1}$, where each block $B_x$ is the $m$-bit binary representation of $\lfloor p_x\cdot 10^l\rfloor\bmod 2^m$ ($\lfloor\cdot\rfloor$ denotes the floor of a number), and $B_x\lVert B_{x+1}$ denotes the concatenation of $B_x$ and $B_{x+1}$.
\end{enumerate}
\section{Statistical Performance Analysis}\label{sec:4}
The proposed scheme is a kind of dedicated hash function, the security of which is hard to prove, and it is commonly evaluated by means of statistical analysis. To facilitate comparison and discussion, we consider two typical instances QHFM-P-264 and QHFM-P-296 of the proposed scheme, where QHFM-P-$L$ produces $L$-bit hash values and will be compared with the existing DQW-based hash functions with $L$-bit output length. The values of $m,l,\theta_0,\theta_1$, and $\alpha$ for the two instances are the same, which are taken to be $8,8,\pi/4,\pi/3$, and $\pi/4$, respectively; the only distinction between QHFM-P-296 and QHFM-P-264 lies in the value of $n$, which are taken to be 37 for the 296-bit output length and 33 for the 26-bit output length.

Our statistical tests considers four kinds of properties: sensitivity of hash value to message, diffusion and confusion properties, uniform distribution property, and collision resistance property, the latter three are assessed by analyzing the same collection of hash values, whose input messages come from the public arXiv dataset (see \underline{https://www.kaggle.com/Cornell-University/arxiv}).
\subsection{Sensitivity of Hash Value to Message}\label{sec:4.1}
Let $msg_0$ be an original message and $msg_i$ ($i\in\{1,2,3\}$) the slightly modified result of $msg_0$, the sensitivity of hash value to message is assessed by comparing the hash value $H(msg_i)$ of $msg_i$ with the hash result $H(msg_0)$ of $msg_0$. Specifically, the original and modified messages are obtained under the following four conditions.
\begin{itemize}
    \item[$\bullet$] Condition 0: Randomly select a record from the dataset, take the texts of the abstract field within this record as $msg_0$;
    \item[$\bullet$] Condition 1: Invert a bit of $msg_0$ at a random position and then get the modified message $msg_1$;
    \item[$\bullet$] Condition 2: Insert a random bit into $msg_0$ at a random position and then obtain $msg_2$;
    \item[$\bullet$] Condition 3: Delete a bit from $msg_0$ at a random position and then obtain $msg_3$.
\end{itemize}
Corresponding to the above conditions, we list, as an example, four hash values in hexadecimal format produced by QWM-P-296 as follows.
\begin{itemize}
    \item[$\bullet$] $H(msg_0)=$``D8 CD B4 F1 A9 A8 E4 F8 60 F7 6F 74 B8 A7 C9 60 E8 7F 53 7F 10 F9 B0 1D 9D D6 37 A5 F1 8E F3 D5 71 5A 28 7B B1'';
    \item[$\bullet$] $H(msg_1)=$``EB C6 F0 5F 74 5B B2 14 72 5F B7 29 CF 7F C6 96 E8 8B 87 8C 9A E9 50 51 8A D5 5D 62 0E ED F8 7B CE 15 D1 7F A7'';
    \item[$\bullet$] $H(msg_2)=$``90 7E 3E 3F 8A D5 D5 6B 70 41 BD F9 37 B2 55 22 2B F1 76 C0 D3 09 E4 6C B5 88 CA C8 07 6D 7B 3F 22 6A 28 BF 21'';
    \item[$\bullet$] $H(msg_3)=$``65 19 6A A5 EE AC 65 A9 52 FA E5 30 B7 22 56 67 51 F2 8F 8E CD B0 6E 6F 04 25 9F 2B 8D E0 97 CE 49 82 66 7A A8''.
\end{itemize}
The plots of $H(msg_0)$, $H(msg_1)$, $H(msg_2)$, and $H(msg_3)$ in binary format are shown in Fig.~\ref{fig:fig1}, where each asterisk (*) in the $j$th subgraph ($j>0$) marks a different bit between $H(msg_{j})$ and $H(msg_0)$. It indicates that a slight modification in the input message can lead to a significant change in the hash result, and the positions of changed bits are evenly distributed over the entire interval [1,296] of position numbers. A similar result can be obtained using QHFM-P with other output lengths, thus the output result of the proposed hash scheme is highly sensitive to its input message.
\Figure[ht](topskip=0pt, botskip=0pt, midskip=0pt){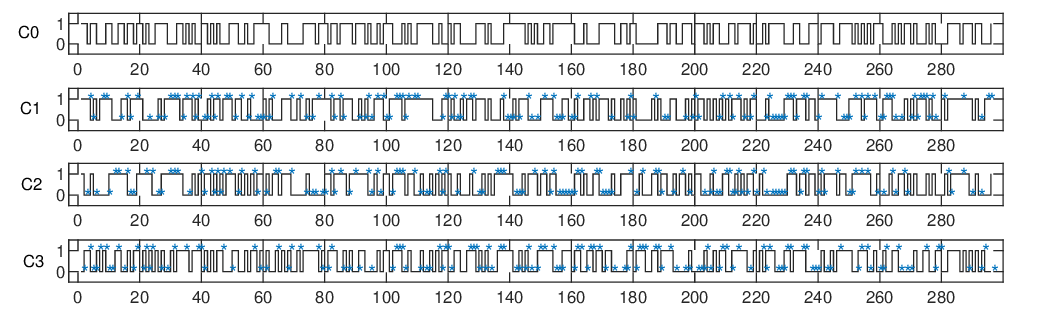}
{Plots of the hash values produced by QHFM-P-296 under the four conditions, where $\text{C}j$ stands for Condition $j$ ($j=0,1,2,3$).
\label{fig:fig1}}
\subsection{Diffusion and Confusion Properties}\label{sec:4.2}
To test the diffusion and confusion properties of the proposed hash function, the statistical experiment getting $msg_0$ and $msg_1$ are independently repeated $N$ times, then the hash values of those $N$ pairs of messages are analyzed. Let $B_i$ ($i\in\{1,\dots,N\}$) be the Hamming distance between $H(msg_0)$ and $H(msg_1)$ obtained in the $i$th experiment, the diffusion and confusion properties are assessed based on the following four indicators:
\begin{enumerate}
    \item[1)] mean changed bit number $\overline{B}=\begin{matrix} \sum_{i=1}^N B_i/N \end{matrix}$;
    \item[2)] mean changed probability $P=\overline{B}/(n\times m)\times 100\%$;
    \item[3)] standard deviation of the changed bit number $\Delta B=\\
    \sqrt{\begin{matrix}\sum_{i=1}^N\left(B_i-\overline{B}\,\right)^2\end{matrix}\big/(N-1)}$;
    \item[4)] standard deviation of the changed probability $\Delta P=\\
    \sqrt{\begin{matrix}\sum_{i=1}^N\left[B_i/(n\times m)-P\right]^2\end{matrix}\big/(N-1)}\times 100\%$.
\end{enumerate}
The ideal value of $P$ is $50\%$, and smaller $\Delta B$ and $\Delta P$ are more desirable. Following Ref.~\cite{Zhou2021Hash}, we take $N=10000$ and use $I_\text{DC}=(\Delta P+|P-50\%|)/2\times 100\%$ as a composite indicator for the diffusion and confusion properties. The test results of the diffusion and confusion properties for the proposed hash functions are presented in Table~\ref{tab:table1}. For comparison, the reported results for the existing DQW-based hash schemes with 296- or 264-bit output length are also listed in Table~\ref{tab:table1}, where Yang21-296 is the second instance (with $p=2/n$) in Ref.~\cite{Yang2021qwHash}.
\begin{table}
    \caption{Test Results of the Diffusion and Confusion Properties}\label{tab:table1}
    \tabcolsep 4.5pt 
    \begin{tabular}{lrrrrr} 
    \hline\noalign{\smallskip}
    \makecell[ml]{Hash Instances\\or Schemes} & $\overline{B}$~~~~~~& $P(\%)$~~& $\Delta B$~~& $\Delta P(\%)$ & $I_\text{DC}(\%)$ \\
    \hline\noalign{\smallskip}
        \rowcolor[gray]{.8}
        QHFM-P-296 & 147.9320 & 49.9770 & 8.5517 & 2.8891 & 1.4560 \\
        QHFM-P-264 & 132.1286 & 50.0487 & 8.1938 & 3.1037 & 1.5762 \\
        \rowcolor[gray]{.8}
        QHFL-296~\cite{Hou2023Hash} & 148.1900 & 50.0600 & 8.5500 & 2.8900 & 1.4750 \\
        QHFL-264~\cite{Hou2023Hash} & 132.0300 & 50.0100 & 8.1100 & 3.0700 & 1.5400 \\
        \rowcolor[gray]{.8}
        QHFM-296~\cite{Zhou2021Hash} & 147.9101 & 49.9696 & 8.5997 & 2.9053 & 1.4679 \\
        QHFM-264~\cite{Zhou2021Hash} & 131.8667 & 49.9495 & 8.1378 & 3.0825 & 1.5665 \\
        \rowcolor[gray]{.8}
        Yang21-296~\cite{Yang2021qwHash} & 147.8640 & 49.9541 & 8.6141 & 2.9102 & 1.4781 \\
        Yang19-264~\cite{Yang2019qwHash} & 131.6803 & 49.8789 & 8.8877 & 3.3666 & 1.7439 \\
        Yang18-264~\cite{Yang2018qwHash264} & 132.1108 & 50.0420 & 8.0405 & 3.0457 & 1.5439 \\
    \hline
    \end{tabular}
\end{table}

It can be seen that the test results for QHFM-P are very close to those for its peers, thus the diffusion and confusion properties of the proposed scheme are on a par with those of existing schemes with output length 296 or 264.
\subsection{Uniform Distribution Analysis}\label{sec:4.3}
The uniform distribution property of the proposed hash function is assessed by analyzing the $N$ pairs of hash values used in the diffusion and confusion test in a different way. Let $T_j$ ($j\in\{1,2,\dots ,n\times m\}$) be the number of experiments in which the $j$th bit of $H(msg_0)$ is different from the $j$th bit of $H(msg_1)$, then the uniform distribution analysis considers the following two indicators:
\begin{enumerate}
    \item[1)] mean number of experiments with flipped hash bit over $n\times m$ bit-positions $\overline{T}=\begin{matrix} \sum_{j=1}^{n\times m} T_j/(n\times m) \end{matrix}$;
    \item[2)] standard deviation of the number of experiments with flipped bit
    $\Delta T=\sqrt{\begin{matrix}\sum_{j=1}^{n\times m}(T_j-\overline{T})^2\end{matrix}\big/(n\times m\!-\!1)}$.
\end{enumerate}
The ideal value of $\overline{T}$ is $N/2$, and smaller $\Delta T$  suggests better uniform distribution property. As shown in Table~\ref{tab:table2}, the values of $\overline{T}$ and $\Delta T$ for the two instances of QHFM-P are close to the corresponding values for the recent DQW-based hash functions\footnote{The values of $\overline{T}$ and $\Delta T$ for QHFL are both not available, so we only compare QHFM with the remaining four recent schemes.} with the same output length. Since the reported value of $N\times P$ (which is equivalent to $\overline{T}$ theoretically, see~\cite{Zhou2021Hash}) for Yang21-296 is 4995.41, the value of $|\overline{T}-5000|$ for Yang21-296 may be considered between (or close to) 1.90 and 4.59, so the $\overline{T}$ value of QHFM-P-296 is on the same level as that of Yang21-296.
\begin{table}
    \caption{Test Results of the Uniform Distribution Property}\label{tab:table2}
    \tabcolsep 4.5pt 
    \begin{tabular}{lrcr} 
    \hline\noalign{\smallskip}
    \makecell[ml]{Hash Instances\\or Schemes} & $\overline{T}$~~~~~~& $\left\vert\overline{T}-5000\right\vert $~~ & $\Delta T$ \\
    \hline\noalign{\smallskip}
        \rowcolor[gray]{.8}
        QHFM-P-296 & 4997.70 & 2.30 & 51.9772 \\
        QHFM-P-264 & 5004.87 & 4.87 & 49.8667 \\
        \rowcolor[gray]{.8}
        QHFM-296~\cite{Zhou2021Hash} & 4996.96 & 3.04 & 48.4334 \\
        QHFM-264~\cite{Zhou2021Hash} & 4994.95 & 5.05 & 48.9253 \\
        \rowcolor[gray]{.8}
        Yang21-296~\cite{Yang2021qwHash} & 4998.10 & 1.90 & N/A \\
        Yang19-264~\cite{Yang2019qwHash} & 4996.60 & 3.40 & N/A \\
        Yang18-264~\cite{Yang2018qwHash264} & 5003.90 & 3.90 & N/A \\
    \hline
    \end{tabular}
\end{table}

\begin{figure}
    \centering
    \includegraphics[width=0.98\columnwidth]{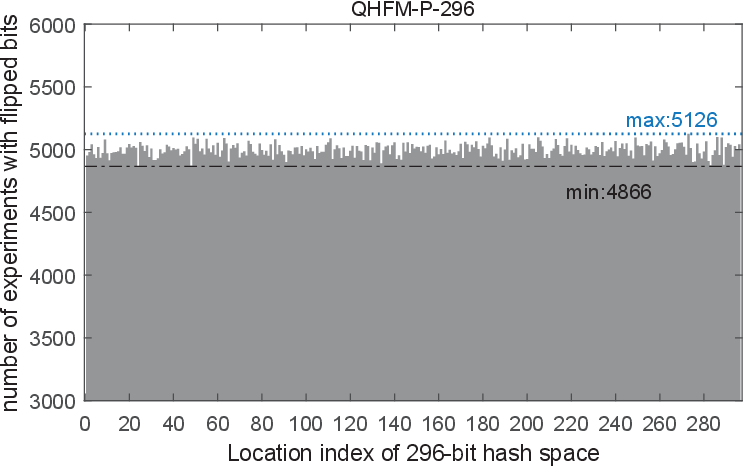}
    \caption{Histogram of the 195-bit hash space, where $N=10000$.}\label{fig:fig2}
\end{figure}

To provide an intuitive description, we plot the number of experiments with flipped hash bit on every bit position of QHFM-P-296 in Fig.~\ref{fig:fig2}, where the number of experiments with flipped hash bit on every bit-position is very close to $N/2$, suggesting that the proposed scheme has good uniform distribution property.
%
\subsection{Collision Resistance}\label{sec:4.4}
The collision resistance test is carried out by counting the number of experiments in which the hash values of the original and modified messages collide at a certain number of bytes, and then comparing the counting result with its theoretical value.

For ease of exposition, we use $\left\{msg_0^{(i)},msg_1^{(i)}\right\}$ to denote the original and modified messages obtained under Condition 0 and Condition 1 in the $i$th experiment, $\left\{H(msg_0^{(i)}),H(msg_1^{(i)})\right\}$ to denote the hash values of $\left\{msg_0^{(i)},msg_1^{(i)}\right\}$, and $g=\lceil (n\times m)/8\rceil $ to denote the number of bytes that a hash result produced by the proposed hash function can be divided into\footnote{If $n\times m$ is not divisible by 8, then add a prefix of $(8-n\times m)\bmod 8$ zeros to the hash value.}. The collision resistance test counts the numbers $\{W_N^e(\omega)|\omega=0,1,\dots,g\}$ of experiments in which $H(msg_0)$ and $H(msg_1)$ have $\omega$ identical bytes ($\omega$ is also called the number of hits). For instance, if the first, third, and fourth bytes of $H(msg_0)$ are respectively the same as the first, third, and fourth bytes of $H(msg_1)$ in the 25th experiment , then $\left\{H(msg_0^{(25)}),H(msg_1^{(25)})\right\}$ makes an incremental contribution of 1 to $W_N^e(3)$.

The theoretical value of $W_N^t(\omega)$ is calculated using the binomial distribution formula
\begin{equation}\label{eq:7}
\begin{split}
    W_N^t(\omega)&=\text{int}\left[N\times P^t(\omega)\right] \\
                 &=\text{int}\left[N\times \frac{g!}{\omega !(g-\omega)!}\left(\frac{1}{2^8}\right)^{\omega} \left(1\!-\!\frac{1}{2^8}\right)^{g-\omega}\right],
\end{split}
                \end{equation}
where $\text{int}\left[\cdot\right]$ denotes rounding a real number to its nearest integer, and $P^t(\omega)$ is the theoretical probability that $\omega$ hits occur in $\{H(msg_0),H(msg_1)\}$. Substituting $g=\lceil 296/8\rceil=37$ and N=10000 into Eq.~(\ref{eq:7}), one can get $\{W_N^t(\omega)|\omega=0,1,2,3\}=\{8652,1255,89,4\}$ for hash functions with 296-bit output length. Similarly, the values of $W_N^t(\omega)$ with $\omega=0,1,2,3$ for 264-bit hash functions are 8788, 1137, 71, and 3, respectively. For both 296- and 264-bit hash functions, the value of $W_N^t(\omega)$ with $\omega\geq 4$ is 0.

Let $P^e(\omega)=W_N^e(\omega)/N$ be the experimental probability that $H(msg_0)$ and $H(msg_1)$ have $\omega$ identical bytes, the collision resistance property of the proposed hash function can be assessed by the Kullback-Leibler divergence (referred to as KL divergence for short) of $P^e$ from $P^t$
\begin{equation}\label{eq:8}
        D_{KL}\left(P^e\lVert P^t\right)=\sum_{\omega = 0}^g P^e(\omega)\text{log}_2\left(\frac{P^e(\omega)}{P^t(\omega)}\right)
\end{equation}
The smaller the value of $D_{KL}\left(P^e\lVert P^t\right)$ is, the closer $P^e$ is to $P^t$. Note that there exist some values of $\omega$ such that $P^e(\omega)=0$ and $P^t(\omega)\neq 0$, hence we cannot use $D_{KL}\left(P^t\lVert P^e\right)$ to indicate the distance between $P^e$ and $P^t$.

In addition to the KL divergence, the mean of the absolute difference per byte between $H(msg_0)$ and $H(msg_1)$ over $N$ independent experiment can also be used to assess the collision resistance property. Let $t(e_j)^{(i)}$ and $t(e'_j)^{(i)}$ be the decimal value of the $j$th byte of $H(msg_0^{(i)})$ and $H(msg_1^{(i)})$, respectively; the mean of the absolute difference per byte is given by
\begin{equation}\label{eq:9}
    \overline{d}_{byte}^{\,e}=\frac{1}{N}\sum_{i = 1}^N\sum_{j=1}^g\frac{1}{g}\left\lvert t(e_j)^{(i)}-t(e'_j)^{(i)}\right\rvert,
\end{equation}
and the theoretical value of $\overline{d}_{byte}^{\,e}$ is $\overline{d}_{byte}^{\,t}=85.33$~\cite{Yang2021qwHash}.

The test results of the collision resistance property for the proposed hash function are shown in Table~\ref{tab:table3}, where $W_N^e(4+)$ denotes the number of experiments in which at least 4 hits occur in $\{H(msg_0),H(msg_1)\}$. The values of $D_{KL}(P^e\| P^t)$ and $\left\lvert \overline{d}_{byte}^{\,e}-\overline{d}_{byte}^{\,t}\right\rvert $ for the two instances of QHFM-P is smaller than or close to those for the existing hash instances with the same output length excepting QHFL, meaning that the collision resistance property of QHFM-P is better than or on a par with that of its peers excepting QHFL. For QHFL-296, the experimental value of $D_{KL}(P^e\| P^t)$ is slightly smaller than that for QHFM-P-296, while the value of $\left\lvert \overline{d}_{byte}^{\,e}-\overline{d}_{byte}^{\,t}\right\rvert$ is greater than that for QHFM-P-296; in similar, the value of $\left\lvert \overline{d}_{byte}^{\,e}-\overline{d}_{byte}^{\,t}\right\rvert$ for QHFL-264 is smaller than that for QHFM-P-264, while the value of $D_{KL}(P^e\| P^t)$ for QHFL-264 is slightly greater than that for QHFM-P-264. Taking into account the fluctuation of the experimental results, the collision resistance property of QHFM-P can be regarded as being on the same level as that of QHFL.
\begin{table*}
    \caption{Test Results of The Collision Resistance property}\label{tab:table3}
    \setlength{\tabcolsep}{8pt} 
    \begin{tabular}{llccc}
    \hline\noalign{\smallskip}
    \textrm{Hash Instances or Schemes} & $\{W_N^{\,e}(\omega)|\omega=0,1,2,3,4+\}$ & $D_{KL}\left(P^e\lVert P^t\right)$ & $\overline{d}_{byte}^{\,e}$ & $\left\lvert\overline{d}^{\,e}_{byte}-\overline{d}^{\,t}_{byte}\right\rvert$ \\
    \hline\noalign{\smallskip}
        \rowcolor[gray]{.8}
        QHFM-P-296 & $\quad 8637,1260,100,3,0$ & 0.0001461 & 85.30 & 0.03 \\
        QHFM-P-264 & $\quad 8794,1143,61,2,0$ & 0.0001513 & 85.40 & 0.07 \\
        \rowcolor[gray]{.8}
        QHFL-296\cite{Hou2023Hash} & $\quad 8637,1278,81,4,0$ & 0.0000998 & N/A & 0.15 \\
        QHFL-264\cite{Hou2023Hash} & $\quad 8784,1133,82,1,0$ & 0.0002427 & N/A & 0.02 \\
        \rowcolor[gray]{.8}
        QHFM-296\cite{Zhou2021Hash} & $\quad 8605,1312,81,2,0$ & 0.0003610 & 85.36 & 0.03 \\
        QHFM-264\cite{Zhou2021Hash} & $\quad 8762,1159,74,5,0$ & 0.0001457 & 85.27 & 0.06 \\
        \rowcolor[gray]{.8}
        Yang21-296~\cite{Yang2021qwHash} & $\quad 8321,1547,110,22,0$ & 0.0086163  & 85.22 & 0.11 \\
        Yang19-264~\cite{Yang2019qwHash} & $\quad 9019,923,52,2,4$ & 0.0056472 & 89.76 & 4.43 \\
        Yang18-264~\cite{Yang2018qwHash264} & $\quad 8904,1026,68,2,0$ & 0.0009686 & 83.64 & 1.69 \\
    \hline\noalign{\smallskip}
    \end{tabular}
\end{table*}
%
\section{Stability with respect to coin parameters}\label{sec:5}
Recall from section~\ref{sec:3} that the proposed hash function is parameterized by three integers and three angles, among which the two coin angles are crucial components of the underlying controlled alternate quantum walks and may affect the four statistical properties considered in section~\ref{sec:4}.

To explore how robust the hashing properties of QHFM-P are with respect to different coin angles, we uniformly divide $(0,\pi/2)$ into $c=30$ subintervals, take the endpoints (except 0 and $\pi/2$) of those subintervals as the candidate values of each coin parameter, and then conduct $N=2048$ experiments for each pair of values. During the experiments for each angle pair, primary indicators of each type of statistical property are calculated. Specifically, we take $P$ together with $\Delta P$, $\overline{T}$ together with $\Delta T$, and $D_{KL}(P^e\| P^t)$ together with $\left\lvert \overline{d}_{byte}^{\,e}-\overline{d}_{byte}^{\,t}\right\rvert$ to be the primary indicators of the diffusion and confusion properties, the uniform distribution property, and the collision resistance property, respectively. Following Ref.~\cite{Hou2023Hash}, we take the mean Jensen-Shannon divergence~\cite{Fuglede2004Diverg} (over $N$ random experiments) between the resulting probability distributions corresponding to the original and modified messages to be the quantitative indicator of the sensitivity of hash value to message. Suppose, in an experiment, the probability distribution produced by the underlying controlled alternate quantum walks controlled by $msg_j$ ($j=0,1,2,3$) is $P_j$, the Jensen-Shannon divergence (referred to as JS divergence for short) between $P_0$ and $P_1$ is 
\begin{equation}\label{eq:10}
D_{JS}(P_0,P_1)=\frac{D_{KL}(P_0\| M)}{2}+\frac{D_{KL}(P_1\| M)}{2}, 
\end{equation}
where $M=(P_0+P_1)/2$, and $D_{KL}(P_0\| M)$ is the KL divergence of $P_0$ from $M$ (see Eq.~(\ref{eq:8})).

\begin{figure*}[ht]
    \centering
    \includegraphics[width=17cm]{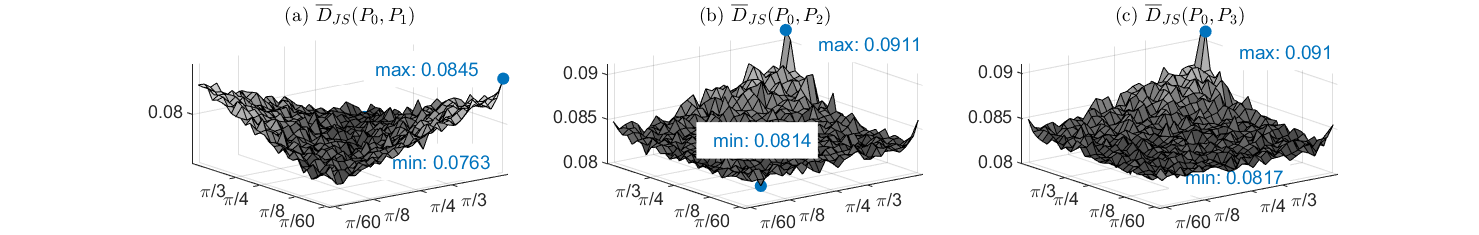}
    \caption{Mean JS divergences between $P_0$ and $P_j$ ($j=1,2,3$) for $\theta_0$ and $\theta_1$ in $[\pi/60,29\pi/60]$.}
    \label{fig:fig3}
\end{figure*}

\begin{figure*}
    \centering
    \includegraphics[width=11cm]{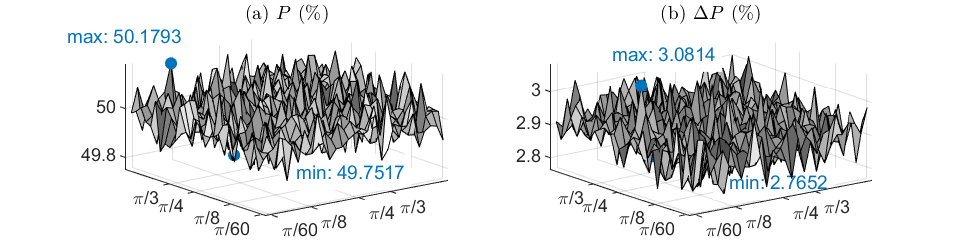}
    \caption{$P$ and $\Delta P$ for $\theta_0$ and $\theta_1$ in $[\pi/60,29\pi/60]$.}\label{fig:fig4}
\end{figure*}

\begin{figure*}
    \centering
    \includegraphics[width=11cm]{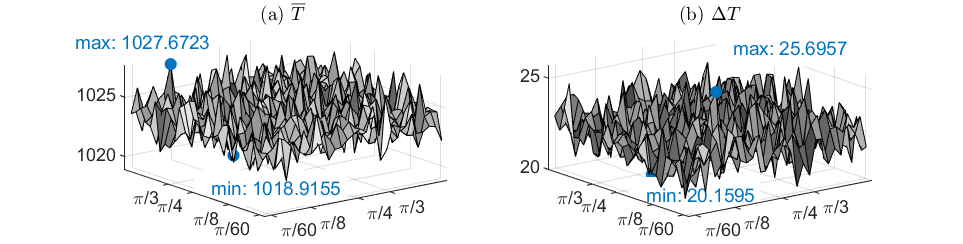}
    \caption{$\overline{T}$ and $\Delta T$ for $\theta_0$ and $\theta_1$ in $[\pi/60,29\pi/60]$.}\label{fig:fig5}
\end{figure*}

\begin{figure*}
    \centering
    \includegraphics[width=11cm]{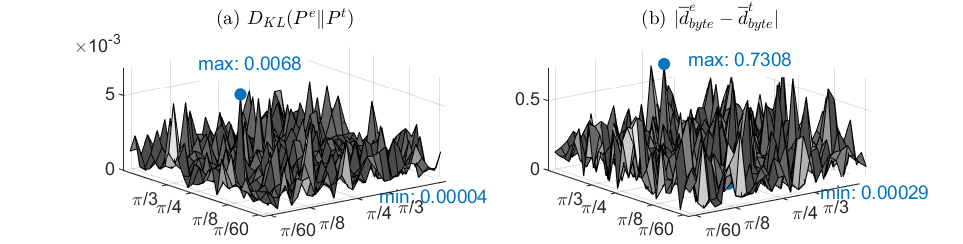}
    \caption{$D_{KL}(P^e\| P^t)$ and $\left\lvert \overline{d}_{byte}^{\,e}-\overline{d}_{byte}^{\,t}\right\rvert$ for $\theta_0$ and $\theta_1$ in $[\pi/60,29\pi/60]$.}\label{fig:fig6}
\end{figure*}

The results of the stability test for the four kinds of properties of QHFM-P-296 are illustrated in Figs.~\ref{fig:fig3} to \ref{fig:fig6}, where $\overline{D}_{JS}(P_0,P_j)$ ($j=1,2,3$) in Fig.~\ref{fig:fig3} denotes the average value of $D_{JS}(P_0,P_j)$ over $N$ experiments. One may notice that the shape of Fig.~\ref{fig:fig4}(a) is identical to that of Fig.~\ref{fig:fig5}(a), this is because $\overline{T}$ is directly proportional to $P$ when the $N$ pairs of original and modified messages used in the diffusion and confusion test are re-used in the uniform distribution test.

It can be seen from Fig.~\ref{fig:fig3}(a) that the average JS divergences between $P_0$ and $P_1$ for different values of coin parameters fall within a narrow range, indicating that the value of $\overline{D}_{JS}(P_0,P_1)$ is quite stable with respect to $\theta_0$ and $\theta_1$. Similarly, the maximum and minimum values of $\overline{D}_{JS}(P_0,P_2)$ (or $\overline{D}_{JS}(P_0,P_3)$) are very close to each other, suggesting that the sensitivity of hash value to message of the proposed hash function is stable with respect to the coin parameters. In Fig.~\ref{fig:fig4}, the mean changed probability $P$ fluctuates around $50\%$; meanwhile, the value of the standard deviation $\Delta P$ is small and does not vary significantly for different values of $\theta_0$ and $\theta_1$. Hence, the diffusion and confusion properties of QHFM-P is robust with regard to different coin angles. Likewise, the experimental values of $\overline{T}$ for different coin angles are close to the theoretical value $N/2$, and the results of $\Delta T$, $D_{KL}(P^e\| P^t)$, and $\left\lvert \overline{d}_{byte}^{\,e}-\overline{d}_{byte}^{\,t}\right\rvert$ displayed in Figs.~\ref{fig:fig5} and \ref{fig:fig6} are relatively small, which suggests that the uniform distribution property and the collision resistance property of the proposed hash function are also stable with respect to coin parameters.
\section{Time and Space Complexity}\label{sec:6}
Similar to existing DQW-based hash schemes, the proposed hash algorithm can be efficiently computed using classical computer. Thus, it is more appropriate to consider QHFM-P as a classical algorithm and to concentrate on classical complexity.

By rewriting the 4-term basis state $\Ket{x,d_2,d_1,c}$ of QWHF-P in $\mathcal{H}_p\otimes\mathcal{H}_{d_2}\otimes\mathcal{H}_{d_1}\otimes\mathcal{H}_c$ as a 2-term basis state $\Ket{x,j}=\Ket{x,2^2d_2+2^1d_1+2^0c}$ in $\mathcal{H}_p\otimes\mathcal{H}^8$, where $\mathcal{H}^8$ is the 8-dimensional Hilbert space, the quantum state of the walker performing parity-dependent quantum walks with one- and two-step memory on a cycle of length $n$ after $t$ time steps can be expressed as 
\begin{equation}\label{eq:11}
    \Ket{\psi_t}=\sum_{x = 0}^{n-1}\sum_{j = 0}^{7}A_t^{x,j}\Ket{x,j}.
\end{equation}
The results of sequentially performing the coin operator, the direction-determine transform, and the shift operator of QW2M-P on the computational states $\Ket{x,j}$ are tabularized in Table 4. For ease of notation, we denote $C^{(1)}$ by $\bigl(\begin{smallmatrix} a_1 & b_1 \\ c_1 & d_1 \end{smallmatrix}\bigr)$ and write $(x\pm 1)\bmod n$ shortly as $x\pm 1$. In the first three columns of Table~\ref{tab:table4}, the second terms $j$ of the basis states and the transformed results are written in binary format, for the binary representation $j_2j_1j_0$ of $j$ indicates the values of $d_2$, $d_1$, and $c$ in a natural way.
\begin{table*}
    \caption{Actions of the coin, the direction-determine, and the complete one-step transforms of QW2M-P}\label{tab:table4}
    \begin{tabular}{lccc}
    \hline\noalign{\smallskip}
    $\Ket{x,j}$ & $I_{4n}\otimes C^{(1)}\Ket{x,j}$ & $(I_n\otimes D^{(1)})(I_{4n}\otimes C^{(1)})\Ket{x,j}$ & $S(I_n\otimes D^{(1)})(I_{4n}\otimes C^{(1)})\Ket{x,j}$ \\
    \hline\noalign{\smallskip}
    $\Ket{x,000}$ & $a_1\Ket{x,000}+c_1\Ket{x,001}$ & $a_1\Ket{x,000}+c_1\Ket{x,011}$ & $a_1\Ket{x-1,0}+c_1\Ket{x+1,3}$ \\
    $\Ket{x,001}$ & $b_1\Ket{x,000}+d_1\Ket{x,001}$ & $b_1\Ket{x,000}+d_1\Ket{x,011}$ & $b_1\Ket{x-1,0}+d_1\Ket{x+1,3}$ \\
    $\Ket{x,010}$ & $a_1\Ket{x,010}+c_1\Ket{x,011}$ & $a_1\Ket{x,110}+c_1\Ket{x,101}$ & $a_1\Ket{x+1,6}+c_1\Ket{x-1,5}$ \\
    $\Ket{x,011}$ & $b_1\Ket{x,010}+d_1\Ket{x,011}$ & $b_1\Ket{x,110}+d_1\Ket{x,101}$ & $b_1\Ket{x+1,6}+d_1\Ket{x-1,5}$ \\
    $\Ket{x,100}$ & $a_1\Ket{x,100}+c_1\Ket{x,101}$ & $a_1\Ket{x,010}+c_1\Ket{x,001}$ & $a_1\Ket{x+1,2}+c_1\Ket{x-1,1}$ \\
    $\Ket{x,101}$ & $b_1\Ket{x,100}+d_1\Ket{x,101}$ & $b_1\Ket{x,010}+d_1\Ket{x,001}$ & $b_1\Ket{x+1,2}+d_1\Ket{x-1,1}$ \\
    $\Ket{x,110}$ & $a_1\Ket{x,110}+c_1\Ket{x,111}$ & $a_1\Ket{x,100}+c_1\Ket{x,111}$ & $a_1\Ket{x-1,4}+c_1\Ket{x+1,7}$ \\
    $\Ket{x,111}$ & $b_1\Ket{x,110}+d_1\Ket{x,111}$ & $b_1\Ket{x,100}+d_1\Ket{x,111}$ & $b_1\Ket{x-1,4}+d_1\Ket{x+1,7}$ \\
    \hline\noalign{\smallskip}
    \end{tabular}
\end{table*}

Combining Eq.~(\ref{eq:11}) with the first and last columns of Table~\ref{tab:table4}, one can obtain the relation between $\{A_{t+1}^{x,j}|x\in\mathbb{Z}_n,j\in\mathbb{Z}_8\}$ and $\{A_t^{x,j}|x\in\mathbb{Z}_n,j\in\mathbb{Z}_8\}$ (hereafter, simply $\{A_{t+1}^{x,j}\}$ and $\{A_{t}^{x,j}\}$, respectively)  after one step of QW2M-P as follows:
\begin{equation*}
    \begin{aligned}
    A_{t+1}^{x,0}=a_1 A_t^{x+1,0}+b_1 A_t^{x+1,1}, \\A_{t+1}^{x,1}=c_1 A_t^{x+1,4}+d_1 A_t^{x+1,5},\\
    A_{t+1}^{x,2}=a_1 A_t^{x-1,4}+b_1 A_t^{x-1,5}, \\A_{t+1}^{x,3}=c_1 A_t^{x-1,0}+d_1 A_t^{x-1,1},\\
    A_{t+1}^{x,4}=a_1 A_t^{x+1,6}+b_1 A_t^{x+1,7}, \\A_{t+1}^{x,5}=c_1 A_t^{x+1,2}+d_1 A_t^{x+1,3},\\
    \end{aligned}
\end{equation*}
\begin{equation}\label{eq:12}
    \begin{aligned}
    A_{t+1}^{x,6}=a_1 A_t^{x-1,2}+b_1 A_t^{x-1,3}, \\A_{t+1}^{x,7}=c_1 A_t^{x-1,6}+d_1 A_t^{x-1,7}.
    \end{aligned}
\end{equation}

Relation~(\ref{eq:12}) shows that the eight amplitudes of being at position $x$ at time $t+1$ can be calculated from the amplitudes of being at positions $x\pm 1$ at time $t$ using 16 multiplications and 8 additions, meaning that the $8n$ amplitudes of being at $n$ locations can be calculated using $O(n)$ basic arithmetic operations. Similarly, one can obtain the relation between $\{A_{t+1}^{x,j}\}$ and $\{A_t^{x,j}\}$ for QW1M-P, where the amplitudes can also be updated using $O(n)$ basic operations at each time step.

If one wants to obtain an $L$-bit hash value ($L$ is a multiple of $m$) of an $M$-bit message, then the walker moves $M$ steps on a cycle with $L/m = O(L)$ nodes, here $m$ is constant with respect to $M$. The assignment of the initial amplitudes $\{A_0^{x,j}\}$ can be carried out with $O(L)$ time and $O(L)$ memory space, the values of $\{A_M^{x,j}\}$ can be calculated from $\{A_0^{x,j}\}$ using $O(ML)$ basic operations with $O(L)$ space, and the hash value can be computed from $\{A_M^{x,j}\}$ using $O(L)$ multiplications and $O(L)$ modulo operations with $O(L)$ space. As a result, the time and space complexity of QHFM-$L$ are $O(ML)$ and $O(L)$, respectively, which are the same as those of the state-of-the-art hash functions~\cite{Zhou2021Hash,Hou2023Hash,Yang2021qwHash,Yang2019qwHash,Yang2018qwHash264} based on discrete quantum walks.
\section{Conclusion}\label{sec:7}
In this article, a new QWM-based hash function QHFM-P is proposed and analyzed. Similar to the existing QWM-based hash function~\cite{Zhou2021Hash}, the proposed scheme is also constructed by using quantum walks with one- and two-step memory; the major distinction lies in that the underlying walks with two-step memory of QHFM and QHFM-P are different extensions of Mc Gettrick's QW1M model~\cite{Gettrick2010QW1M}. The proposed hash function has the same time and space complexity as those of QHFM, and it also has near-ideal statistical properties. It is noteworthy that the four kinds of statistical properties of QHFM-P are quite stable with respect to the coin angles, which suggests the robustness of the hashing properties of the proposed scheme.

In the future, we will try to identify the region of stability in the plane of coin and other parameters, examine the stability with respect to various direction-determine transforms, and establish the conditions for constructing good QWM-based hash functions.
%

\EOD

\end{document}